# OBSERVING GRAVITATIONAL LENSES FROM INTENSITY FLUCTUATIONS: EXPERIMENTAL VALIDATION OF THE TECHNIQUE

Ermanno F. Borra

Centre d'Optique, Photonique et Laser,

Département de Physique, Université Laval, Québec, Qc, Canada G1K 7P4

(email: borra@phy.ulaval.ca)

**SHORT TITLE:** Observing gravitational lenses from intensity fluctuations

**ABSTRACT**
It has been proposed to study gravitational lenses from measurements of the spectrum of the fluctuations of the output current of a quadratic detector. The spatial correlation coefficient of the source is the fundamental parameter of the technique. The experimental work discussed in this article confirms that the correlation coefficient must be evaluated at the frequencies of the spectrum of the current fluctuations. This validates a powerful yet simple technique to find unresolved gravitational lenses and to study the lensing event and the source. The validation is needed before starting the extensive theoretical and observational work that must now follow.

**KEY WORDS:** gravitational lensing – methods: observational – techniques: interferometric

## 1. INTRODUCTION

The discovery of the first gravitational lens by Walsh et al, (1979) has spurred great interest in gravitational lensing and its applications (Blandford & Narayan 1992, Treu 2010). Besides the lensing phenomenon itself, gravitational lensing is of interest because it opens up the possibility of studying the structure of the universe and of finding and studying objects that would otherwise be undetected. Gravitational lenses introduce time delays between the components of the multiple images that they generate (Blandford & Naryan 1992). The time delay work carried out so far has involved measurements of flux time variations within resolved images of strong gravitational lenses. This limits one to long time delays in strongly lensed sources that have intrinsic flux variations. Because a gravitational lens can be modeled by a Young interferometer, interference effects can also be used to study lensing. For example, Spillar (1993) has suggested using radio autocorrelation techniques to study time delays in microlenses.

Borra (1997) has suggested to measure time delays from the spectrum of intensity fluctuations measured with a quadratic detector. Although the technique could be very useful, for example by greatly extending the range of observable time delays, Borra (1997) ended pessimistically by concluding that the spatial extendedness of an astronomical source would make the effect difficult to detect. Borra (2008) reexamined the work of Borra (1997), concluding that Borra (1997) was unduly pessimistic for he assumed that the spatial coherence function, which quantifies the extendedness of the source, had to be computed at the frequency of observation. Borra (2008) uses a theoretical analysis based on statistical optics to show that the coherence function has to be evaluated at the much larger beat frequencies at which the measurements are actually carried out (e.g. hundreds of MHz for a source observed at 1 micron). This would render the technique far more useful.

The work of Borra (2008) is purely theoretical and, as it standard practice in science, experimental confirmation of theoretical work is needed to validate it. This is particularly important in Astronomy since validation of a new experimental technique in the laboratory is highly desirable before one starts time-consuming observational work. Considering the cost of operating an observatory and the limited observing time, one should expect that time allocation committees would be reluctant to grant telescope time on the basis of theoretical work that has not been validated in the laboratory. In this paper, we shall discuss experimental work that confirms the theoretical analysis in Borra (2008).

## 2. INTENSITY FLUCTUATIONS FROM GRAVITATIONAL LENSING AND THE COHERENCE FUNCTION

The basic theory has been discussed by Borra (1997), Borra (2008) and references therein. We summarize here, for convenience, the sections relevant to the discussion in this article.

The usual geometrical model for the two components of a gravitational lens (Refsdal 1964; Press & Gunn 1973) models it by a Young interferometer that introduces time delays between two interfering beams. In fringe interferometry one observes intensity fringes. While in fringe interferometry the visibility of the fringes disappears if the optical path difference between the interfering beams exceeds the coherence length, interfering beams that have very large optical path differences generate a recombined beam that has a spectral distribution modulated with periodic minima and maxima. Spectral modulation occurs for optical path differences that far exceed the coherence length of the interfering beams. This surprising statement is experimentally confirmed and has firm theoretical bases (Mandel 1962). Alford & Gold (1958) found spectral modulation for an optical path difference of 64.2 m, far exceeding the coherence length of their source (an electrical spark). They detected the spectral modulation in the beat spectrum measured in the output current of a quadratic detectors (a photomultiplier). Cielo, Brochu & Delisle (1975) detected modulation directly in the primary spectrum of a white light source observed through a Michelson interferometer giving an optical path difference of 300 meters, orders of magnitudes greater than the coherence length.

In the discussion that follows we distinguish between the primary spectrum and the beat spectrum. The primary spectrum is the spectrum of the source (e.g. the visible spectrum) before it is measured by a quadratic detector (e.g. a photomultiplier). The beat spectrum is the spectrum of the fluctuations of the output current of the quadratic detector. It is called a beat spectrum because it is generated by the electromagnetic waves of the primary spectrum that beat among each other (Givens 1961). Mandel (1962) gives a full theoretical justification of the Alford & Gold effect, while Givens (1961) gives a less rigorous but easier to follow physical explanation. Borra (1997) and Borra (2008) discuss mostly the theory in Givens (1961); however below we shall use the theory in Mandel (1962) because it is more appropriate to the discussion of the experimental data. Mandel (1962) uses statistical optics theory to show that the spectral density of the primary spectrum of the combined beams is given by

$$\Phi_{33}(\nu) = \Phi_{11}(\nu)\{1 + [2(\overline{I_1}\overline{I_2})^{1/2}/(\overline{I_1} + \overline{I_2})]\gamma_{12}(0)\cos(2\pi\nu\tau)\}, \qquad (1)$$

where $\Phi_{11}(\nu)$ is the normalized input spectral density, $\tau$ the time delay, $\gamma_{12}(0)$ the normalized cross-correlation function at $\tau=0$ and $\nu$ the frequency of light. The terms $\overline{I_1}$ and $\overline{I_1}$ are the intensities in the 2 beams of the interferometer and the bar above the intensities signifies the ensemble average. The coefficient $\gamma_{12}(0)$ is obtained from the normalized cross-correlation of the electric field from the 2 interfering beams $\gamma_{12}(\tau)$ after using the usual assumption used in fringe interferometry (Klein & Furtak 1986) that

$$\gamma_{12}(\tau) = \gamma_{12}(0)\gamma_{11}(\tau) \quad . \qquad (2)$$

Equation 1 takes into account spatial coherence effects caused by the extendedness of the source with the $\gamma_{12}(0)$ factor. The coefficient $\gamma_{12}(0)$ is the same function that is used in fringe interferometry to quantify the contrast in the intensity interference fringes given by a Young interferometer (Klein & Furtak 1986) and is thus also given by the Fourier transform of the intensity distribution across the source. It must be evaluated at a specific frequency. In fringe interferometry one uses the middle frequency of the bandpass of the spectral distribution of the filter placed before the detector.

Mandel (1962) then proceeds to derive the spectral modulation of the output current

$$\psi_{44}(v') = \psi^I_{44}(v')\{1 + [2(\overline{I_1 I_2})/(\overline{I_1} + \overline{I_2})]\gamma^2_{12}(0)\cos(2\pi v'\tau)\} , \qquad (3)$$

where $\psi^I_{44}(v')$ is the spectral distribution for $\tau = 0$. Equation 3 shows that the spectral distribution $\psi_{44}(v')$ at $\tau >> 0$ is modulated by a $\cos(2\pi v'\tau)$ term.

With the separation of the $\gamma_{12}(\tau)$ in two factors (equation 2) Mandel (1962) makes the critical assumption that spectral purity is conserved. Spectral purity signifies that the spectral distribution of the two incoming beams and the combined beams are the same (Mandel 1961). While this is true for fringe interferometry, it is obviously not true for the Alford & Gold effect since equation 1 shows that the two interfering beams have the spectrum given by $\Phi_{11}(v)$, while the combined beams have the spectrum given by $\Phi_{33}(v)$. Equation 1 clearly shows that $\Phi_{33}(v) \neq \Phi_{11}(v)$ since the $\cos(2\pi v\tau)$ term modulates $\Phi_{11}(v)$. Note that Mandel (1962) does indeed state at the end of p. 1336 that the experiment does not conserve spectral purity. Furthermore Mandel (1961) gives the Alford & Gold experiment as an example of non-conservation of spectral purity. Presumably he made this assumption, although he stated it was not correct, simply because it allowed him to pursue a purely analytical treatment. Furthermore, this may not affect the main conclusions of his paper which is not concerned with the evaluation of $\gamma_{12}(0)$.

The basic problem that we will address is concerned with the evaluation of the spatial coherence function $\gamma_{12}(0)$. It is given the Fourier transform of the intensity distribution across the source and the Fourier integral must be evaluated at a specific frequency. Borra (1997) assumed that the integral has to be evaluated at the frequency $v$ in the primary spectrum; while Borra (2008) argues that it has to be evaluated at the beat frequency of measurement $v'$.

## 3. EXPERIMENTS

Experimental work has been carried out to verify the conclusions of the theoretical analysis in Borra (2008). The goal of the experiments is to test whether the basic conclusion in Borra (2008): that a microscopic source can only beat with itself is correct. The important practical consequence of this is that, to obtain the beat spectrum of an extended source, one must add the contributions of individual beat spectra originating in small areas of the source.

Fig. 1 shows the optical set up used to obtain the experimental data. The light sources are Erbium-doped fiber amplifiers emitting at a wavelength of 1550 nm with a bandpass width of 50 nm. The two coupled Mach-Zehnder interferometers are each made of two single-mode optical fibers having an index of refraction *n = 1.45* and lengths differences giving an Optical Path Difference (OPD) of 3.12 meters for one and 6.35 meters for the other one. The outputs from the two interferometers are then fed into a single fiber and measured by a quadratic detector. The output currents of the quadratic detector, an InGaAs photodiode having an output electric bandpass of 350 MHz, are measured with a sampling scope at a sampling frequency of 20 GHz. The digital output of the scope is then used to obtain, with Matlab software, the autocorrelation function

$$I \otimes I = \int_{-\infty}^{\infty} I(t+t')I(t')dt' \quad . \tag{4}$$

Finally, the power spectrum is obtained from a Fourier transform of the autocorrelation function

Fig. 2 shows the autocorrelation functions, in arbitrary units, obtained for the output of the interferometer having an OPD of 3.12 meters alone, for the output of the interferometer having an OPD of 6.35 meters alone and for the output of the combined interferometers. The sharp signals peaking at ± 3.12-m and at ± 6.35 m respectively carry the signatures of the 2 path differences. In Fig. 2 one cannot distinguish the signals at ± 3.12-m and ± 6.35-m in the combined outputs of the interferometers from those in the individual ones because they are superposed and they have the same shapes and amplitudes. There is no significant degradation in the combined signals with respect to the individual ones. This fact is crucial to understand the conclusions of this article.

Fig. 3 shows the power spectra, in arbitrary units, obtained from the Fourier transform of the autocorrelation functions shown in Figure 2. It also gives, for comparison purposes, the numerical sum of the power spectra for the 3.12-meters and the 6.35-meters OPDs.

## 4. DISCUSSION

We shall divide our discussion below in 2 parts following the treatment in Borra (2008). Firstly, if the hypothesis in Borra (2008) is correct, in the spectrum of the current fluctuations of the combined interferometers, beats should occur only among waves within the spectrum of a single interferometer and there should not be any beats between the spectra of the two interferometers. Secondly, as a consequence of this hypothesis, to take into account the effect of extendedness on the spectral contrast in the beat spectrum of an extended source one must add the individual beat spectra from small sub-regions of the source.

**4.1 Beating only occurs among waves within the spectrum of a single interferometer.**
To understand the phenomenon responsible for the spectral modulation of the current fluctuations, consider that the current fluctuations are generated by wave beats among the frequencies $\nu$ in the primary spectrum (Givens 1961). The beat spectrum (equation 3) is modulated because of the modulation of the primary spectrum (equation 1) that gives minima and maxima in the spectral density.

Let us consider first how the spectral modulation in the primary spectrum of the interferometer having an OPD of 3.12-m gives the spectral modulation shown in Fig. 3. Consider a maximum in its primary spectrum occurring at a frequency $\nu_0$. The wave at frequency $\nu_0$ will beat with a wave at frequency at $\nu_0 + \delta\nu$, with $\delta\nu \approx 0$. Because the spectral densities at $\nu_0$ and $\nu_0 + \delta\nu$ are essentially equal to each other and at the maximum value of the spectral density, there will be a maximum in the beat spectrum at a beat frequency $\nu' = (\nu_0 + \delta\nu) - \nu_0 = \delta\nu \approx 0$. This maximum at $\nu' = 0$ can be seen in the 3.12-m OPD beat spectrum in Fig. 3. Consider now the beating of the wave at frequency $\nu_0$ with a wave at frequency $\nu_0 + 33.13$ MHz, which is a frequency at which (equation 1) there is a minimum in the spectral density of the primary spectrum: There will obviously be a minimum in the beat spectrum at $v' = 33.13$ MHz. This minimum can be seen in the 3.12-m OPD beat spectrum in Fig. 3. Continuing the same reasoning we predict maxima at $v' = m\ 66.26$ MHz, with $m$ an integer number, and minima at $v' = 33.13 + m\ 66.26$ MHz in the beat spectrum. These minima and maxima can be seen in the 3.12-m OPD spectrum in Fig.3.

A similar treatment can be done with the interferometer having an OPD of 6.35 meters. However the maxima in the beat spectrum now occur at $v' = m\ 32.6$ MHz and the minima at $v' = 16.3 + m\ 32.6$ MHz. They can be seen in the 6.35-m OPD spectrum in Fig.3.

Let us now consider the hypothetical beating between the combined spectra from the two interferometers. An analysis similar to the one carried out in the two previous paragraphs can be carried out. Note however that because the ratio between the two optical path differences is 2.035, and therefore $\neq 2$, the difference between the frequencies of the maxima of the 2 sources in the primary spectrum varies with the frequency of the primary spectrum. The shift is similar to the shift with increasing frequency seen in Fig. 3 between the maxima of the 3.12-m OPD beat spectrum and the maxima of the 6.35-m OPD beat spectrum. Because the location of a maximum in the beat spectrum is given by that frequency difference, the maxima (and minima) of the beat spectra will vary within a 66 MHz range. Consequently, the expected effect caused by the beating among the 2 spectra is to decrease the contrast in the beat spectrum of the combined interferometers with respect to the numerical sum spectrum shown in Figure 3.

To better understand this, consider now a wave at a maximum in the primary spectrum of the interferometer having an OPD of 3.12-m and let us assume that it beats with the spectrum of the interferometer having an OPD of 6.35-m. Let us first consider a particular small spectral region of the primary spectrum where the spectra of both interferometers have each a maximum at the same frequency $v_1$. If the two spectra having different OPDs beat together, the maximum at the resulting beat frequency will be at $v' = 0$. The same maximum at $v_1$ in the interferometer having an OPD of 3.12-m will also beat with the remainder of the spectrum of the interferometer having an OPD of 6.35-m giving maxima at $v' = m\ 66.26$ MHz and minima at $v' = 33.13 + m\ 66.26$ MHz. Let us now consider another region of the primary spectrum where the spectrum of the interferometer having an OPD of 3.12-m has a maximum at a frequency $v_2$ and the spectrum of the interferometer having an OPD of 6.35-m now has a minimum at the same frequency $v_2$. Obviously beating will now give a minimum at a frequency $v' = 0$ MHz in the beat spectrum and the first maximum will then occur a $v' = 33.13$. There will then be other maxima at $v' = 33.13 + m\ 66.26$ MHz and other minima at $m\ 66.26$ MHz. The maxima in this case correspond to the minima in the previous case, where both spectra had a maximum at the same frequency $v_1$. Obviously, locations of the maxima will vary between 0 and 66.3 MHz in the regions of the primary spectrum such that $v1 < v < v2$; as well as in spectral regions outside of that frequency interval. This will give the blurring mentioned in the previous paragraph.

The spectrum of the combined beams in Fig. 3 does not show a decrease in contrast with respect to the sum. To the contrary, Fig. 3 shows that the spectrum of the combined interferometer has the same modulation seen in the numerical sum. The data therefore show that the spectra from the 2 interferometers do not beat with each other, thereby confirming this conclusion in Borra (2008).

Note that there is a continuum underneath the modulation of the power spectrum which is twice as strong in the combined spectrum than in the summed spectrum. Because this continuum comes from the Fourier transform of the sharp peak of the autocorrelation function centered at an OPD = 0, this is to be expected since the autocorrelation function (equation 4) depends on the square of the combined signal $I = I_1+I_2$: Obviously, since $(I_1 + I_2)^2 \approx 2(I_1^2 + I_2^2)$ for $I_1 \approx I_2$, this explains the factor of two in the continuum strengths in Fig. 3. On the other hand, the spectral modulation comes from the Fourier transform of the peaks centered at ± 3.12-m and ± 6.35-m. This can be shown from the shift theorem of Fourier theory using an analytical treatment similar to the one used by Givens (1961) and Borra (2008). They use it for a time-dependent pulse $V(t)$, having a frequency spectrum given by the Fourier transform of $V(t)$, sent into an interferometer with an optical path difference $c\tau$. The Fourier transform of two pulses shifted by $\pm \tau/2$ yields a frequency spectrum modulated by a $\cos(2\pi\nu\tau/2)$ factor. In our case we also have 2 peaks in the autocorrelation function separated by times given by the OPD. Because these peaks are the same in the autocorrelations of the single interferometers and in the autocorrelations of the combined interferometers, their spectral modulations, obtained by Fourier transform, are also the same. This same modulated spectrum is added to the aforementioned continuum by the integral of the Fourier transform,

**4.2 To take into account the extendedness of a source one must add individual beat spectra.**

The data shown in Fig. 3 also confirm the validity of the analysis carried out in section 3.2 of Borra (2008) to quantify the spectral contrast in the beat spectrum. Borra (2008) starts from an equation that models the beat spectrum from a region of an extended source, small enough that spatial coherence effects are negligible, by

$$D(\nu') = S(\theta)\Psi(\nu')[1+\cos(2\nu\pi\nu'\tau)], \qquad (5)$$

where $\psi(\nu')$ is the spectral density of the unmodulated beat spectrum and $S(\theta)$ is the intensity in the source at angle $\theta$. Equation 5 approximates well the shape of the beat spectrum that is actually observed (Alford & Gold 1958, Basano & Ottonello 2000) as well as the spectra of the 6.35-m OPD and 3.12-m OPD in Fig. 2. It is also equivalent to equation 3, which uses the notation in Mandel (1962), for $I_1 = I_2$ (the two interfering beams have the same intensity) and $\gamma^2_{12}(0) = 1$ (a source small enough that spatial coherence effects are negligible).

To take extendedness into account, one must add the contributions of different sub-regions of the source, each having a different optical path difference $\tau$. Borra (2008) argues that, since independent sources do not beat among themselves, to take into account the effect of extendedness on the beat spectrum, one should add individual beat spectra. Consequently, what degrades the spectral modulation of the beat spectrum from an extended macroscopic source is the superposition of the beat spectra coming from its many small sub-regions having negligible coherence effects. Borra (2008) therefore integrates, after some algebraic manipulations, equation 5 over the surface of the source to obtain the equation that quantifies the spectral contrast in the beat spectrum

$$D(v') = S_0 \psi(v')[1 + \gamma(a,v')\cos(2\pi v'\tau)], \tag{6}$$

which is, with a slightly different notation, equivalent to equation 3 for $I_1 = I_2$. Equation 6 is valid only if the combined beat spectrum from the superposition of beams having different optical path differences is given by the sum of the individual beat spectra. The parameter that quantifies the contrast is

$$\gamma(a,v') = \mathrm{Re}[\int_{-\infty}^{\infty} i(\theta)e^{-i2\pi v' a\theta/c} d\theta], \tag{7}$$

where Re[] signifies that we take the real part of the complex integral, $a$ is the impact parameter of the gravitational lens and $i(\theta)$ is the normalized angular distribution function $S(\theta)$ in equation 5. The impact parameter $a$ quantifies the time delays among the two images of a lensed source, depends on the mass of the lensing object, and is equivalent to the separation between the slits of a Young interferometer (Press & Gunn 1973). The function $\gamma(a,v')$ is the Fourier transform of the normalized intensity distribution $i(\theta)$. The important feature of equation 7 that makes the proposed technique useful, is that $\gamma(a,v')$ is a function of the beat frequency $v'$.

Figure 3 shows that if one combines two beams having different optical path differences, the combined beat spectrum is given by the sum of the individual beat spectra. Consider that a different optical path is what distinguishes two different sub-regions of a lensed source and that the detector sees the sum of the contributions from these sources: Figure 3 thus confirms that to take extendedness into account one should add the individual beat spectra, each modulated by a cosine term $[1+\cos(2v\pi v'\tau_i)]$ depending on the optical path difference $\tau_i$ of each individual source $i$ to obtain the combined beta spectrum.

In conclusion, by showing that the spectrum of the combined beams is the same as the sum of the spectra of the individual beams having different optical path differences, the experiments confirm that the assumptions used by Borra (2008) to derive equations 6 and 7 are correct: The effect of extendedness on the spectral contrast in the beat spectrum of an extended source is to add individual beat spectra each having a different optical path difference $\tau$.

## 5. CONCLUSION

The experiments presented in section 3 confirm that the theoretical conclusion reached by Borra (2008) that a microscopic source can only beat with itself is correct. Consider next that to take into account the extendedness of a source observed through a gravitational lens, modeled by a Young interferometer, one must add the contributions of all the different regions of the source, each having a different optical path difference. Fig. 3 shows that the beat spectrum of the combined beams of 2 interferometers having different optical path differences is modulated by the sum of two cosine functions having different periods respectively set by the optical path differences in the individual interferometers. It therefore also validates the conclusion in Borra (2008) that to quantify the effect of extendedness on the spectrum of a source one must add the individual beat spectra of each sub-region of the source.

The experimental results therefore validate a powerful novel technique to find unresolved gravitational lenses and to study the lensing event and the source. Measurements of the spectrum of the fluctuations of the output current of the quadratic detector of a telescope can be used to find unresolved astronomical gravitational lenses and determine time delays that can be used for astronomical studies.

This experimental validation of the theory in Borra (2008) is needed before starting the extensive theoretical and observational work that should now follow. Borra (1997) and Borra (2008) assume the simple model, needed for an analytical discussion, of two equal-intensity interfering beams. In practice, we should expect more complicated situations. For example, the lens could produce multiple beams of unequal intensities. Also, there may be multiple lensing objects in the observed beam. This may for example be the case for a distant quasar lensed by an intervening galaxy. Theoretical modeling will be needed to understand this, akin to the extensive theoretical modeling that has been done for the light curves generated by microlensing. The experimental validation is also needed prior to starting observational work since observational follow-up will require telescope time on large telescopes. Large telescopes are needed because of the degeneracy parameter considerations discussed in Borra (1997) and Borra (2008), Time allocation committees of large telescopes are unlikely to grant telescope time without prior experimental confirmation of the theory.

The utilization in section 3 of the autocorrelation function (Equation 4) to obtain the power spectrum in section 3 demonstrates a simple technique to observe gravitational lenses. This technique is interesting because all one needs from observations at the telescope is the digitized output of the current from a quadratic detector. Quadratic detectors are standard telescope detectors. It is therefore easy to implement. In section 3 we used a commercially available oscilloscope to digitize the output current from the detector. Presumably, the same instrument could be used at the telescope. The digital technique has however the inconvenient that one must store large quantities of data, a problem that could be alleviated, for long delay times, by using appropriately long sampling intervals. For very short time delays, it may be preferable to measure the power spectrum directly with appropriate electronics as done by Alford & Gold (1958). The electronic technique works quite well for short delay times and generates less data but is less versatile and more difficult to implement.

The interest of the spectral modulation techniques comes from several features that are discussed at length in Borra (1997) and Borra (2008). Borra (2008) briefly discusses some applications of the technique. For example, it could be used to find clumps of dark matter, find information on the lensing event and the source itself. Perhaps the most interesting application is to measure time delays to determine cosmological parameters (Refsdal 1964, Blandford & Naryan 1992, Treu 2010). While time delays have been used for such a purpose by measuring variations of luminosity in the lensed source, they are difficult to measure. They would be far easier to measure by observing the beat spectrum of the source.

**ACKNOWLEDGEMENTS**

This research has been supported by the Natural Sciences and Engineering Research Council of Canada. The data discussed in section 3 were obtained by J.-D. Deschenes.

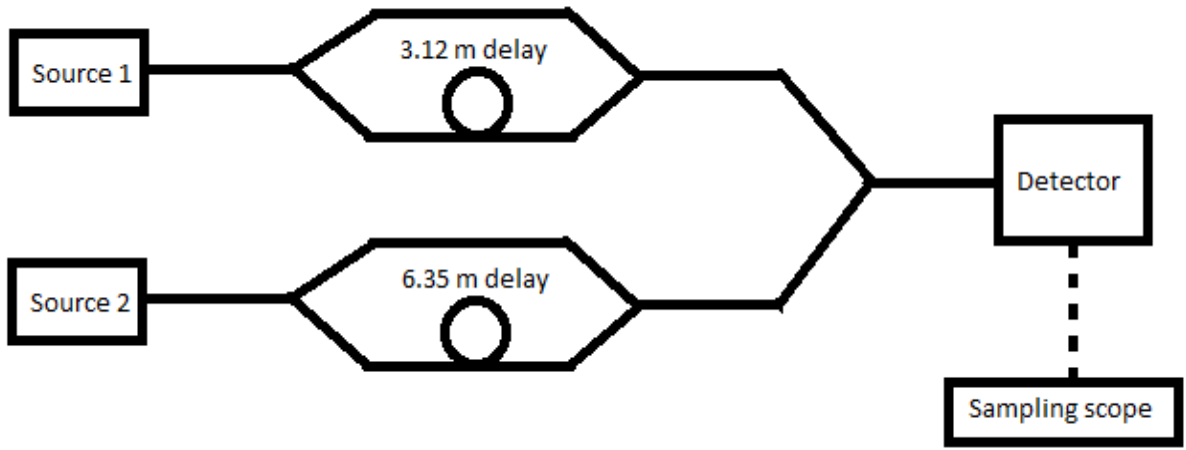

Figure 1 It shows the optical set up used to obtain the experimental data. The light sources are Erbium-doped fiber amplifiers emitting at a wavelength of 1550 nm with a bandpass width of 50 nm. The two coupled Mach-Zehnder interferometers are each made of two optical fibers having an index of refraction of 1.45 and optical path differences of 3.12 meters for one and 6.35 meters for the other one. The signals from the two interferometers are then fed into a single fiber and finally observed by a quadratic InGaAs detector having an electric bandpass 350 MHz. The fluctuations of the output currents of the detector are measured with a sampling scope.

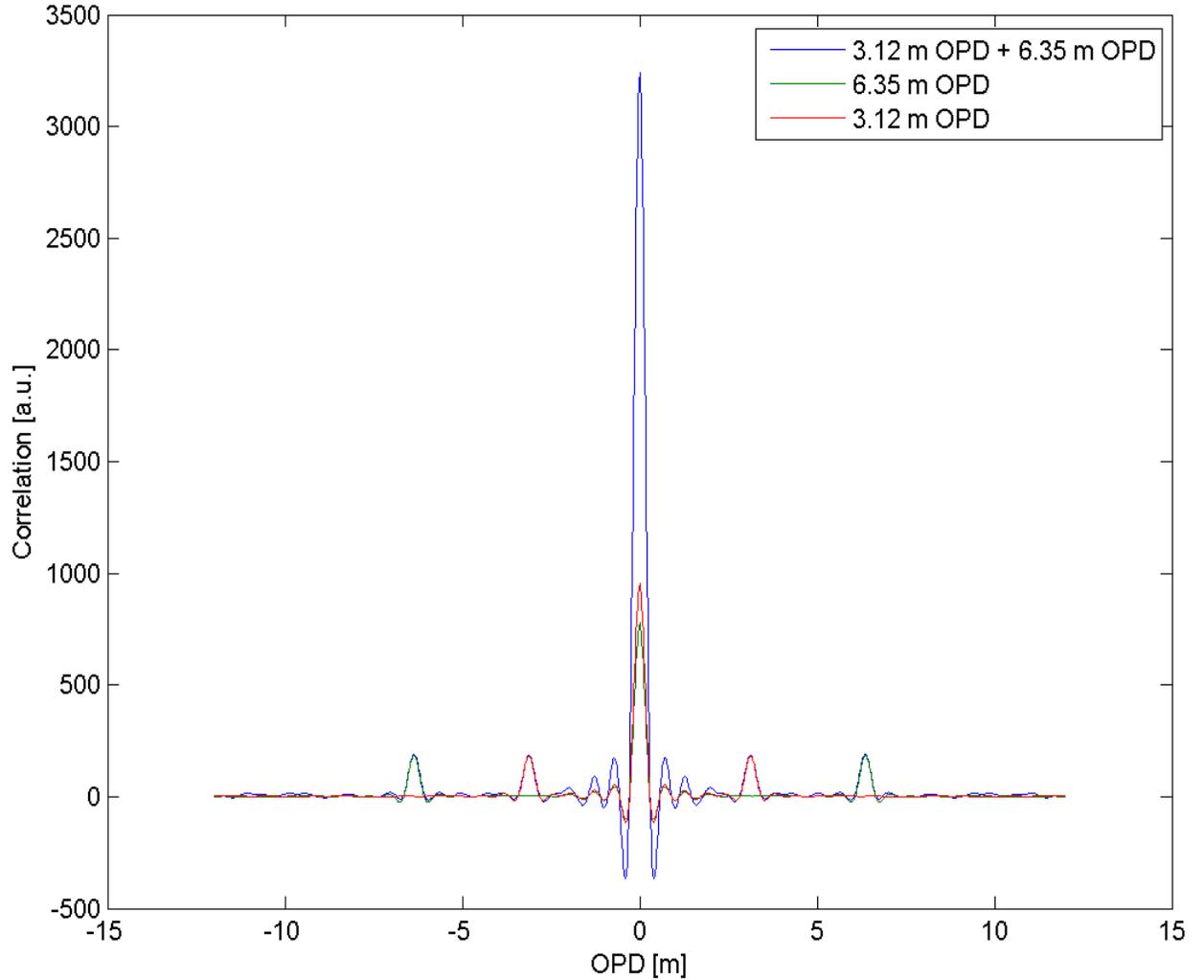

Figure 2 It shows the three autocorrelation functions, in arbitrary units, obtained for the interferometer having an OPD of 3.12 meters alone, for the interferometer having an OPD of 6.35 meters alone and for the output of the combined interferometers. The signals at ± 3.12 meters and at ± 6.35 meters have the same shapes and amplitudes in the combined signal and in the separate signals from the two interferometers alone. This is why one cannot distinguish them in the figure. There is no degradation in the combined signals with respect to the individual ones.

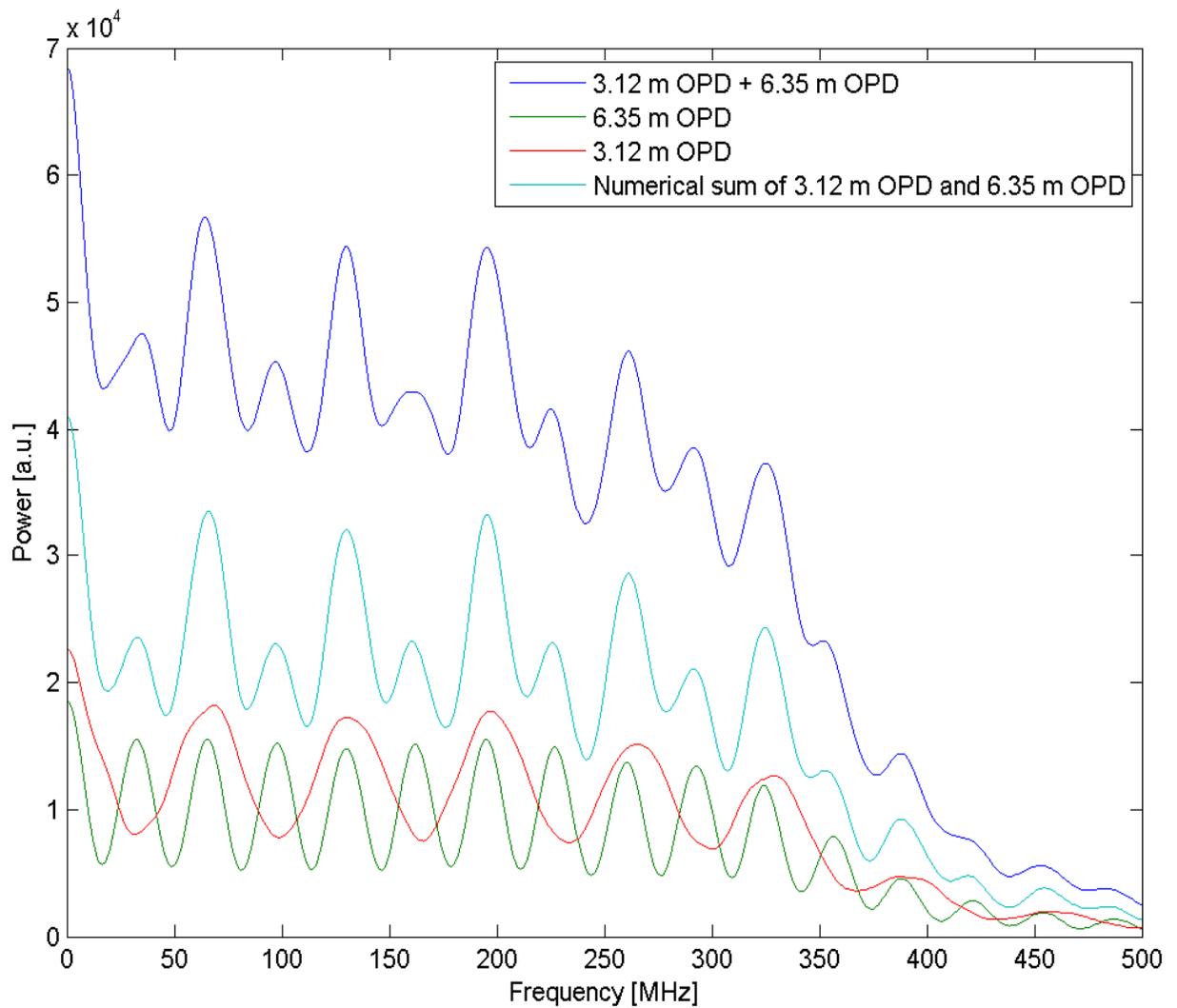

Figure 3 It shows the power spectra, in arbitrary units, obtained from the Fourier transform of the autocorrelation functions shown in Fig. 2. It also gives, for comparison purposes, the numerical sum of the power spectra for the 3.12-meters and the 6.35-meters OPDs